\documentclass[twocolumn,showpacs,preprintnumbers,amsmath,amssymb]{revtex4}
\usepackage{graphicx}

\begin{document}
{\it \small Accepted for publication in Physical Review Letters}

\title{Experimental analysis of the Strato-rotational Instability \\in a cylindrical Couette flow}
\author{M. Le Bars and P. Le Gal}

\affiliation{Institut de Recherche sur les Ph\'enom\`enes Hors Equilibre,
\\UMR 6594, CNRS \& Aix-Marseille Universit\'e, 49 rue
F. Joliot-Curie, BP146, 13384 Marseille C\'edex 13}

\date{\today}

\begin{abstract}

This study is devoted to the experimental analysis of the
Strato-rotational Instability (SRI). This instability affects the
classical cylindrical Couette flow when the fluid is stably
stratified in the axial direction. In agreement with recent
theoretical and numerical analyses, we describe for the first time
in detail the destabilization of the stratified flow below the
Rayleigh line (i.e. the stability threshold without
stratification). We confirm that the unstable modes of the SRI are
non axisymmetric, oscillatory, and take place as soon as the
azimuthal linear velocity decreases along the radial direction.
This new instability is relevant for accretion disks.

\end{abstract}

\maketitle

Taylor-Couette flow (or cylindrical Couette flow) is certainly one
of the most popular laboratory flows and its study has already led
to an abundant scientific literature. Several articles review the
theoretical and experimental features of this flow that consists
of a simple shear between two
 co-axial rotating cylinders \cite{diprima}.  Its linear stability is given by
 the Rayleigh criterion which states that rotating shear flows are stable when their
angular momentum increases radially. The richness of the
transition diagram that reports the different flow patterns and
their subsequent transitions to chaos and turbulence has been
extensively explored by Andereck et al. \cite{andereck}.
Amazingly, the literature about Taylor-Couette flows in the
presence of an axial stable stratification is much more limited
even though it finds direct applications in geophysics
(atmospheric or oceanic flows) or astrophysics (accretion disks
for instance). The first experiment on stratified Taylor-Couette
flow was performed by Thorpe \cite{thorpe} in 1966. His results
were further extended by the experiments of Withjack and Chen
\cite{withjack}, Boubnov et al. \cite{boubnov95}, Caton et al.
\cite{caton} and by the numerical simulations of Hua et al.
\cite{hua}. Most of these studies were performed with a stationary
outer cylinder and all concluded that stable vertical
stratification stabilizes the flow and reduces the axial
wavelength of the unstable modes
which bifurcate through Hopf bifurcations.\\
\begin{figure}[h!]
\begin{center}
\includegraphics[width=8cm]{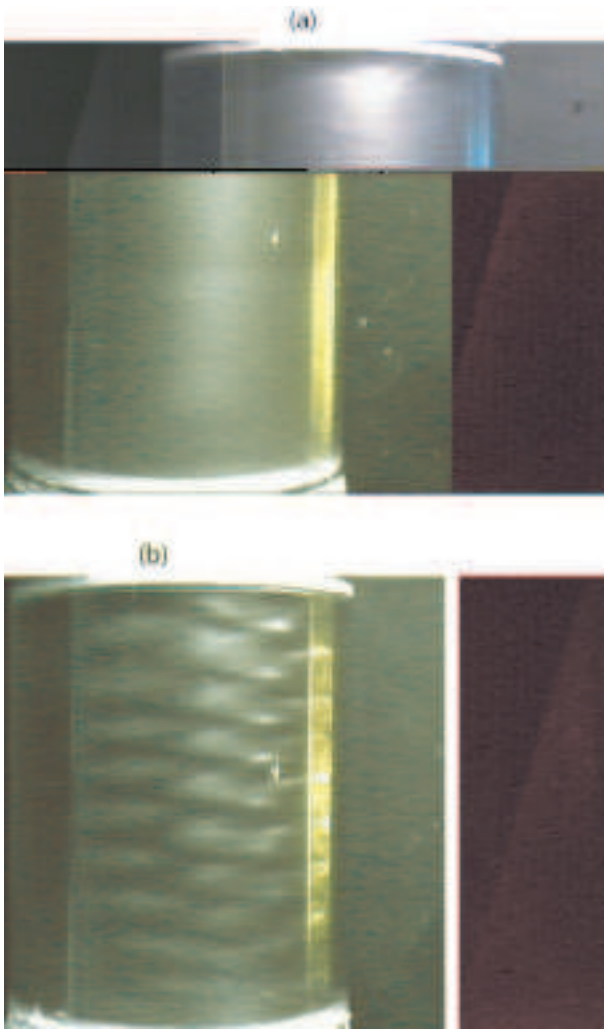} \hfill
\caption{Kalliroscope visualization of the stable Couette flow in
pure water (a) and of the SRI in the vertically stratified salt
water with $Fr=0.5$ (b) at the same Reynolds number $R_e= 1155$
($\Omega_o= 1.07 rad/s$, $\Omega_i= 1.5 rad/s$). The braid pattern
of the SRI is formed by the superimposition of two identical
helicoidal waves (azimuthal wavenumber $m=4$) propagating in
opposite vertical directions. The laser sheet shows the meridional
section of the travelling pattern (on the right part of the
device).} \label{visu}
\end{center}
\end{figure}
In astrophysics, Keplerian flows whose angular velocity is
$\Omega(r)\sim r ^{-3/2}$, are always stable to infinitely small
perturbations in respect to the Rayleigh criterion \cite{dub}. As
a consequence of this stability, several scenarios have been
recently proposed to justify the existence of turbulence in
accretion disk flows. Ingredients other than pure shear then need
to be taken into account in order to trigger instabilities and
transition to turbulence in these flows. One of the most popular
scenarios relies on the Magneto-Rotational Instability
\cite{balbus} but other instabilities such as the elliptical
instability \cite{lebo} or non linear shear instabilities
\cite{dub} have also been suggested.\\
In 2001, Molemaker et al. \cite{molemakerPRL} and Yavneh et al.
\cite{yavnehJFM} predicted that cylindrical Couette flows in a
stratified fluid may become unstable even if the Rayleigh
criterion for stability was verified, i.e. in the corresponding
stable regime of pure fluid flow. Moreover, and contrary to the
classical Taylor vortices of the centrifugal instability, the most
unstable modes should be non-axisymmetric. This theoretical
analysis has then been continued in an astrophysical context by
Shalybkov and R\"udiger \cite{rudi} and extended to the stability
of accretion disk Keplerian flows by Dubrulle et al.
\cite{dubrulle} who gave to this new instability its now accepted
appellation Strato-Rotational Instability (SRI). To the best of
our knowledge, and despite explicit calls recently published
\cite{molemakerPRL,rudi}, no experimental validation of these
theoretical predictions has yet been provided, apart from a short
comment made by Withjack and Chen \cite{withjack} in 1974 and a
stability curve that was determined by Boubnov and Hopfinger in
1997 \cite{boubnov}. We present here the first quantitative
experimental evidences of the Strato-rotational Instability and
its expected helicoidal modes.\\
Our study is based on Kalliroscope visualizations performed in a
classical Couette device with a ratio of inner to outer cylinder
diameters equal to $\eta = 0.80$. The shear created by the
differential rotation of these cylinders is measured by the ratio
$\mu$ of the outer cylinder rotation speed to the inner cylinder
rotation speed: $\mu = \frac{\Omega_o}{\Omega_i}$. Using $\mu$ and
$\eta$, the two traditional control parameters for the cylindrical
Couette flow, the Rayleigh criterion reads $\mu = \eta^2$ which is
represented by the Rayleigh line in the ($\Omega_o$,$\Omega_i$)
plane. A salt stratification is realized along the axis of
rotation of the Couette apparatus using the classical
"double-bucket" filling method \cite{Oster}. The corresponding
Froude number of the fluid is chosen equal to 0.5 which is the
estimated value for accretion disks \cite{dubrulle}. Contrary to
the extensive study of Caton et al. \cite{caton} where the outer
cylinder is at rest, the particularity of our study comes from the
possibility to rotate separately both cylinders and therefore to
study the development of non axisymmetric helicoidal modes in
usually stable regions (see fig. 1). As we will see, our
experimental determination of the Strato-rotational Instability
threshold is in excellent agreement with the value $\mu = \eta$
proposed in \cite{rudi} and corresponding to a constant
azimuthal linear velocity along the radial direction.\\
Our Couette device consists of two co-axial cylinders whose length
is $168 mm$. The inner radius of the outer cylinder is $R_o= 69mm$
and the radius of the inner cylinder is $R_i= 55 mm$, conferring a
$14 mm$ gap between the two concentric cylinders. The ratio of the
inner to outer radius of the fluid cavity is equal to $\eta = 0.8$
explicitly chosen to allow some comparisons with the results of
Shalybkov and R\"udiger \cite{rudi}. The cylinders are positioned
vertically. The inner cylinder is made of black polished acetal
plastic and the outer one is of transparent glass so
visualizations can be performed with the help of Kalliroscope
flakes. To complement these bulk visualizations that use a
classical light bulb for illumination, a laser sheet can also be
installed to get a description of the hydrodynamical structure in
a vertical meridional plane (see figure 1). The cylinders are
driven by a d.c. servo-motor. Rotational speeds are measured by an
optical encoder with an accuracy better than $1\%$. The top and
bottom lids of the device are fixed to the external cylinder and
rotate with it. The experimental protocol is the following. The
classical double-bucket technique \cite{Oster} is used to obtain a
vertical stratification of salt water. The vertical density
gradient is measured before each run with the help of a density
meter or a conductivity meter using a calibrated plunging probe. A
linear gradient is quite easily achieved and its value leads to
the determination of the Brunt-V{\"{a}}is{\"{a}}l{\"{a}}
pulsation: $ N=(-\frac{g}{\rho}~\frac{\partial \rho }{\partial
z})^{1/2}$ with an accuracy better than 5 $\%$. When changing the
salt concentration in one of the filling bucket, $N$ can be chosen
between $0$ and $3.1 rad/s$. The Froude number $Fr$ is then
calculated by the ratio of the inner cylinder rotation speed
$\Omega_i$ to $N$. We have fixed {\`a priori} its value to
$Fr=0.5$ in order to compare our experimental results with the
numerical predictions of \cite{rudi}. Once the density gradient is
established, the inner cylinder rotation speed $\Omega_i$ is thus
fixed for each run, which also fixes the value of the Reynolds
number: $R_{e}= \frac{\Omega_{i}~R_{i}~(R_{o}-R_{i})}{\nu}$, where
$\nu$ is the viscosity of water. For each run, the initial
condition is solid body rotation obtained when both cylinders
rotate at the same speed, then we systematically change the
angular velocity ratio $\mu = \frac{\Omega_o}{\Omega_i}$ by slowly
decreasing $\Omega_o$. When the threshold for the SRI is reached a
single travelling helicoidal wave appears in the flow. After some
minutes, a vertical counter propagative wave, having identical
azimuthal, temporal and vertical (in absolute value) frequencies
is superimposed on the first one, and creates a braid pattern as
illustrated in figure 1-b). We believe that the initial
dissymmetry between both waves comes from initial conditions and
that the reflection on the lids of the first wave restore the
symmetry between $+k$ and $-k$ waves expected from the theory.
Figure 1-a) shows the pure water cylindrical Couette flow in the
exact same conditions and demonstrates the stability of the flow
in the absence of stratification. Let us note that whereas it is
known that stratification damps vertical motions and instabilities
in some cases \cite{thorpe,boubnov,hua,caton}, it is quite clear
here that it
can also destabilize shear flows.\\
\begin{figure}[h]
\begin{center}
\includegraphics[width=8cm]{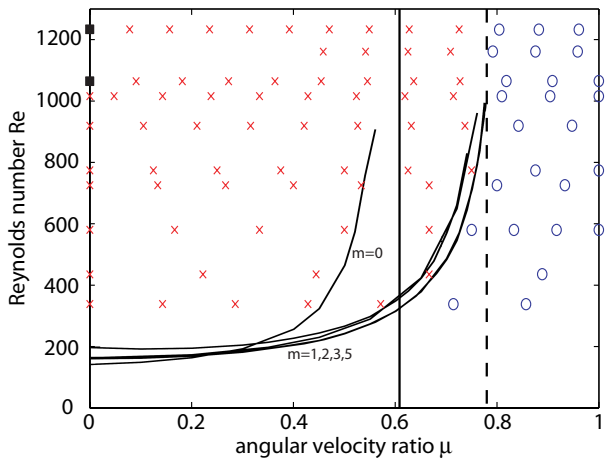}\hfill
\caption{Stability diagram of the Strato-Rotational Instability in
the ($R_e$,$\mu$) parameter plane for a Froude number of $0.5$
obtained for N ranging from $0.85 rad/s$ to $ 3.1 rad/s$. $\circ$
and $\times$ stand respectively for experimental observations of
stable and unstable flows. Squares stand for stationary
axisymetric Taylor vortices. These experimental data points are
superimposed on the solid lines of the theoretical stability
diagram of Shalybkov and R\"udiger \cite{rudi} for azimuthal
wavenumbers $m=0$ (e.g. the classical Taylor vortices) and
$m=1,2,3,5$ for $\eta=0.78$ (some helicoidal modes of the SRI).
Vertical lines correspond to the theoretical inviscid limits with
respectively the Rayleigh line $\mu = \eta ^2$ relevant for the
unstratified case (solid line) and the threshold for the
stratified case $\mu = \eta$ as suggested in \cite{rudi} (dotted
line).} \label{seuil}
\end{center}
\end{figure}
Ten series of experiments corresponding to ten different Reynolds
numbers ranging from $339$ to $1210$ have been performed. Figure 2
sumarizes the experimental stability diagram of the flow in the
parameter plane ($\mu$,$R_e$). It shows our experimental data
points ($\times$ for instability, $\circ$ for stable flow)
superimposed on the stability diagram calculated in \cite{rudi}.
As can be observed an excellent agreement is obtained. In
particular, we confirm that the inviscid threshold, proposed by
Shalybkov and R\"udiger \cite{rudi}, is given by $\mu = \eta$.
Further decreasing $\mu$, patterns become increasingly complex as
additional helicoidal modes ($m \neq 0$) are progressively added.
We suggest that the helicoidal modes calculated by Molemaker et
al. \cite{molemakerPRL} and Shalybkov and R\"udiger \cite{rudi}
and observed in our experiments should be by continuity the
"vortex modes" described by Caton et al. \cite{caton} when the
outer cylinder does not rotate. We never found the "standing wave"
regime of Caton et al. \cite{caton} (i.e. their pulsating
axisymmetric modes) within the explored range, presumably because
its unstable region is too narrow and our experimental device is
not precise enough to excite this very marginal state (see the
comparison between the results of \cite{boubnov95} and
\cite{caton}). Nevertheless, in experiments at $\mu = 0$ and large
Reynolds number, we observed the reappearance of stationary
axisymmetric Taylor vortices in agreement with the regime diagram
of Boubnov et al. \cite{boubnov95}. This stationary axisymmetric
mode was never excited when decreasing $\mu$ for $\mu>0$, but once
excited at $\mu =0$, they could then be maintained up to $\mu =
0.2$. These interesting non-linear effects are beyond the scope of
the present paper devoted to the SRI and will be further studied.
\begin{figure}[h]
\begin{center}
\includegraphics[height=11cm]{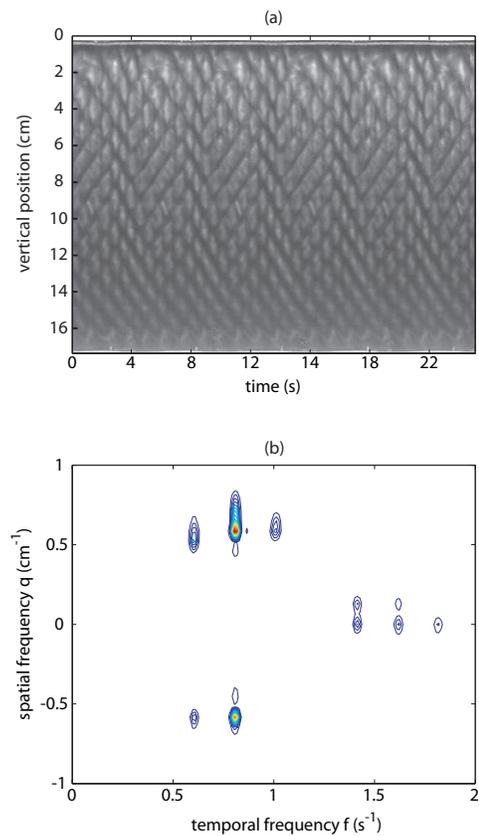} \hfill
\caption{(a) Space-time diagram at threshold of the growing
helicoidal mode presented in figure 1-b), with $Fr=0.5$ and $R_e=
1155$. (b) 2D-Fourier Power spectra of the space-time diagram:
contours are equally space from $0$ to the maximum energy
$E^{max}$ with a spacing of $E^{max}/20$. Non linear interaction
between both helicoidal waves is visible as peaks around $q \simeq
0$ and $f \simeq 1.6 s^{-1}$.} \label{spacetime}
\end{center}
\vspace{-1cm}
\end{figure}
Figure 3-a) shows the space-time diagram of the braid pattern
observed at SRI threshold and shown in figure 1-b). These diagrams
are built by assembling vertical lines from successive video
images, and allow to extract in a systematic way the temporal
frequency $f$ and the axial frequency $q$ of the hydrodynamic
structures using a 2D-Fourier transform, as shown in figure 3-b).
At instability threshold, we systematically recover two
symmetrical points $(f,q)$ and $(f,-q)$ corresponding to the two
helicoidal modes forming the braid pattern. Further away from the
threshold, the 2D-Fourier transforms are more complex and indicate
other peaks, but in the following we only focus on the most
energetic of them. As suggested by the analytical study of
Shalybkov and R\"udiger \cite{rudi}, we did not observe any
systematic variation of the wavenumber $k = 2
\pi~q~(R_{i}(R_{o}-R_{i}))^{1/2}$: over all our experiments, we
measure a mean value of $10.6$ with a standard variation of $1.5$
in close agreement with analytical and numerical results
\cite{yavnehJFM,rudi}. This value corresponds to a wavelength
equal to $1.17$ time the gap, smaller than the standard value in
the absence of stratification (i.e. twice the gap) and confirms
the reduction of the vertical extension of the structures because
of stratification.
\begin{figure}[h]
\begin{center}
\includegraphics[height=11cm]{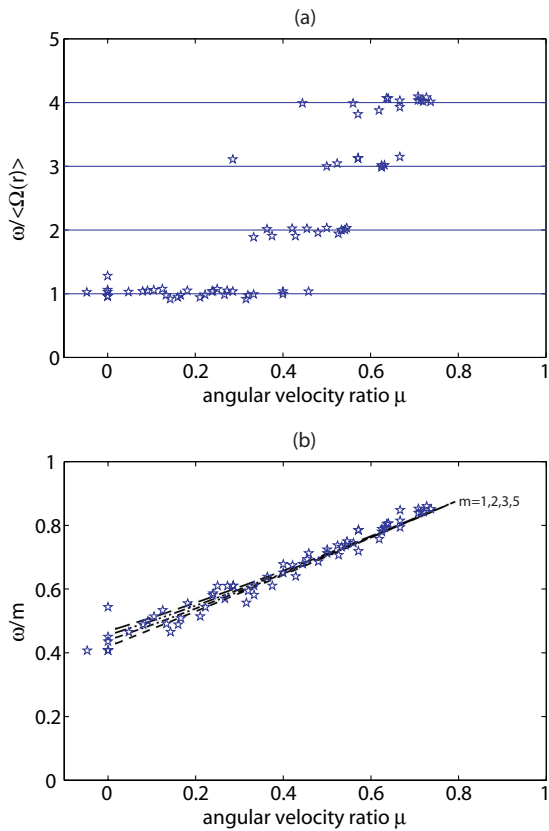}
\caption{(a) Evolution of the pulsation $\omega$ of the prevalent
helicoidal mode divided by the mean angular pulsation
$<\Omega(r)>$ for $Fr=0.5$ obtained for N ranging from $0.85
rad/s$ to $ 3.1 rad/s$: all ratios align with integer values as
expected from the azimuthal periodicity in the cylindrical
geometry. These are highlighted by horizontal lines and correspond
to the azimuthal wavenumber $m$ (from $1$ to $4$) of the selected
mode. These values are used in (b) to rescale $\omega$. All
experimental data points correctly aligns with the analytical
predictions (dotted lines) of the SRI \cite{rudi}.} \label{freq}
\end{center}
\vspace{-1cm}
\end{figure}
 Figure 4-a) presents the evolution of the
experimentally determined pulsation $\omega = 2 \pi f/ \Omega_i$
divided by the mean normalized angular velocity, as a function of
$\mu$. As can be observed, the data points are clustered along
horizontal lines corresponding to integer values. Indeed,
extending the simple advection argument given by Caton et al.
\cite{caton} to $\mu>0$, one can estimate that the helicoidal
patterns are simply advected at the mean angular velocity
$<\Omega(r)>$ of the fluid between the cylinders. Hence the ratio
$\omega$/$<\Omega(r)>$ actually indicates the azimuthal wavenumber
$m$ of the selected mode, ranging from $m=1$ to $m=4$. We confirm
these conclusions by directly measuring the azimuthal wavenumber
at threshold when there is a single helix in the flow. In figure
4-b), we then compare our experimental results using the
previously determined $m$ with the
numerical results of \cite{rudi}. Again, perfect agreement is found.\\
In conclusion, we have extended previous experimental studies of
stratified Taylor-Couette flow by a systematic exploration of the
stable range according to the Rayleigh criterion. We have observed
the new Strato-rotational Instability in the form of
non-axisymmetric oscillating modes. Our results validate the
theoretical results of Molemaker et al. \cite{molemakerPRL},
Yavneh et al. \cite{yavnehJFM}, Shalybkov and R\"udiger
\cite{rudi} and Dubrulle et al. \cite{dubrulle}. In particular,
instability thresholds, axial and azimuthal wavenumbers and
temporal frequencies of the helicoidal travelling modes ($m\neq
0$) have been measured for a fixed Froude number $Fr =0.5$, an
aspect ratio $\eta = 0.80$ and for different Reynolds numbers.
Comparisons with theoretical predictions \cite{rudi} are
excellent. Moreover, let us emphasis that in the inviscid limit,
the SRI threshold tends to the value $\mu = \eta$, contrary to
$\mu = \eta^{2}$ given by the Rayleigh criterion valid in the
unstratified case. As already remarked by Shalybkov and R\"udiger
\cite{rudi}, this implies that stratified Keplerian flows such as
those found in accretion disks and characterized by $\Omega(r)\sim
r ^{-3/2}$ (i.e. $\mu_{Kepler}=\eta^{3/2}$) are stable following
the Rayleigh criterion (i.e. $\mu_{Kepler} > \eta^{2}$), but can
be destabilized by the SRI as $\mu_{Kepler}<\eta$.\\
{\bf Acknowledgements:\\} {The authors are grateful to J. Caillet,
M. Borel and M. Vial-Sablier for their help during the
experiments.\vspace{-7 mm}}

\end{document}